\documentclass[11pt]{article}
\usepackage{graphicx,graphics,floatflt,amssymb,epsf,rotate}
\usepackage{color}
\def\gtap{\raisebox{-.4ex}{\rlap{$\sim$}} \raisebox{.4ex}{$>$}}

\begin{document}
\begin{flushright}
SINP/TNP/2009/06
\end{flushright}
\vskip 30pt

\begin{center}
  {\Large \bf Probing warped extra dimension via ${gg \to h}$ and ${h \to
      \gamma \gamma}$ at LHC } \\
  \vspace*{1cm} \renewcommand{\thefootnote}{\fnsymbol{footnote}} { {\sf Gautam
      Bhattacharyya} and {\sf Tirtha Sankar Ray}
  } \\
  \vspace{10pt} {\small {\em Saha Institute of Nuclear Physics, 1/AF Bidhan
      Nagar, Kolkata 700064, India}}

\normalsize
\end{center}

\begin{abstract}

  The processes $gg \to h$ and $h \to \gamma \gamma$ are of paramount
  importance in the context of Higgs search at the LHC.  These processes are
  loop driven and hence could be sensitive to the presence of any new colored
  fermion states having a large coupling with the Higgs. Such a scenario
  arises in a warped extra dimensional theory, where the Higgs is confined to
  the TeV brane and the hierarchy of fermion masses is addressed by localizing
  them at different positions in the bulk. We show that the Yukawa coupling of
  the Higgs with the fermion Kaluza-Klein (KK) states can be order one
  irrespective of their zero mode masses. We observe that the $gg \to h$ and
  $h \to \gamma \gamma$ rates are substantially altered if the KK states lie
  within the reach of LHC.  We provide both intuitive and numerical comparison
  between the RS and UED scenarios as regards their quantitative impact in
  such processes.

\vskip 5pt \noindent
\texttt{PACS Nos:~ 04.50.Cd, 11.10.Kk } \\
\texttt{Key Words:~~Higgs boson, Warped Extra Dimension}
\end{abstract}

\renewcommand{\thesection}{\Roman{section}}
\setcounter{footnote}{0}
\renewcommand{\thefootnote}{\arabic{footnote}}

{\bf Introduction}:~ For an intermediate mass ($<$ 150 GeV) Higgs boson, the
relevance of its production at the CERN Large Hadron Collider (LHC) via gluon
fusion ($gg \to h$) and its subsequent decay into two photons ($h \to \gamma
\gamma$) cannot be over-emphasized. Since these are loop induced processes, a
natural question arises as how sensitive these processes are to the existence
of new physics.  In this paper, we explore such a possibility by embedding the
Standard Model (SM) in a Randall-Sundrum (RS) warped geometry
\cite{Randall:1999ee}, where the bulk is a slice of Anti-de Sitter space
(AdS$_5$) accessible to some or all SM particles
\cite{bulksm,Gherghetta:2000qt}. The virtues of such a scenario include a
resolution of the gauge hierarchy problem caused by the warp factor
\cite{Randall:1999ee}, and an explanation of the hierarchy of fermion masses
by their respective localizations in the bulk keeping the Higgs confined at
the TeV brane \cite{Huber:2000ie}. Besides, the smallness of the neutrino
masses could be explained \cite{Grossman:1999ra}, and light KK states would
lead to interesting signals at LHC \cite{rslhc}. We demonstrate in this paper
that the loop contribution of the KK towers of quarks and gauge bosons
emerging from the compactification would have a sizable numerical impact on
the $gg \to h$ and $h \to \gamma \gamma$ rates. This happens because the Higgs
coupling to a pair of KK fermion-antifermion is not suppressed by the zero
mode fermion mass and can easily be order one \cite{Bhattacharyya:2008mn}. The
underlying reason is simple.  Although the zero mode wave-functions of
different flavors have varying overlap at the TeV brane depending on the zero
mode masses, the KK profiles of all fermions have a significant presence at
the TeV brane where the Higgs resides. As a result, the KK Yukawa couplings of
different flavors are not only all large, they are also roughly
universal. This large universal Yukawa coupling in the RS scenario constitutes
the corner-stone of our study.  On the contrary, in flat Universal Extra
Dimension (UED) only the KK top Yukawa coupling is large, others being
suppressed by the respective zero mode fermion masses.  We provide comparative
plots to demonstrate how the warping in RS fares against the flatness of UED
for the processes under consideration.

{\bf Warped extra dimension}:~ The extra coordinate $y$ is compactified on an
$S^1/\mathbb{Z}_2$ orbifold of radius $R$, with $-\pi R\leq y\leq\pi R$. Two
3-branes reside at the orbifold fixed points at $y = (0,\pi R)$. The
space-time between the two branes is a slice of AdS$_5$ geometry, and the 5d
metric is given by \cite{Randall:1999ee},
\begin{equation}
\label{metric}
        ds^2=e^{-2\sigma}\eta_{\mu\nu}dx^\mu dx^\nu+dy^2\,
~~\mbox{where}~~\sigma=k|y|~.
\end{equation}
Above, $1/k$ is the AdS$_5$ curvature radius, and $\eta_{\mu\nu}={\rm
  diag}(-1,1,1,1)$.  The natural mass scale associated with the $y=0$ brane
is the Planck scale ($M_P$), while the effective mass scale associated with the
$y=\pi R$ brane is $M_P e^{-\pi kR}$, which is of the order of a TeV for
$kR\simeq 12$. To address the fermion mass hierarchy, the Higgs boson has to
be confined to the TeV brane, thus solving the gauge hierarchy problem in the
same stroke. The bulk contains the fermions and gauge bosons.  After
integrating out the $y$-dependence, the 4d Lagrangian can be written in terms
of the zero modes and their KK towers.  A generic 5d field can be decomposed
as (only fermions and gauge bosons are relevant for us)
\cite{Gherghetta:2000qt}
\begin{equation}
\label{Kaluza-Klein}
        \Phi(x^\mu,y)={1\over\sqrt{2\pi R}}\sum_{n=0}^\infty 
\Phi^{(n)}(x^\mu)f_n(y) , ~~{\rm where}~~
 f_n(y)=\frac{e^{s\sigma/2}}{N_n}\left[J_\alpha(\frac{m_n}{k}e^{\sigma})
     +b_{\alpha}(m_n)\, Y_\alpha(\frac{m_n}{k}e^{\sigma})\right]\, ,
\end{equation}
with $s = (1,2)$ for $\Phi=\{e^{-2\sigma}\Psi_{L,R},A_\mu\}$.  Above,
\begin{equation}
\label{balpha}
    b_\alpha = -\frac{(-r+\frac{s}{2})J_\alpha(\frac{m_n}{k})+\frac{m_n}{k}
     J'_\alpha(\frac{m_n}{k})}{(-r+\frac{s}{2})Y_\alpha(\frac{m_n}{k})
     +\frac{m_n}{k}Y'_\alpha(\frac{m_n}{k})}~,~~\mbox{and}~~ 
N_n \simeq \frac{1}{\sqrt{\pi^2 R~m_n~e^{-\pi kR}}}~,
\end{equation}
where $r=(\mp c,0)$ and $\alpha=(c\pm \frac{1}{2}, 1)$ for $\mathbb{Z}_2$
even/odd modes.  By imposing boundary conditions on $f_n(y)$ in
Eq.~(\ref{Kaluza-Klein}) and in the limit $m_n \ll k$ and $kR \gg 1$, one
obtains the KK masses as (for $n=1,2,\dots$),
\begin{eqnarray}
\label{evenmn}
  m_n\simeq \left(n+\frac{1}{2}|c-\frac{1}{2}| -\frac{1}{4}\right)\pi
     k~e^{-\pi kR}~ \mbox{(fermions)} ~~;~~ 
m_n\simeq \left(n -\frac{1}{4}\right)\pi
     k~e^{-\pi kR}~ \mbox{(gauge bosons)} ~. 
\end{eqnarray}
Now, the Yukawa part of the action with two 5d Dirac fermions
$\Psi_{iL}(x,y)$ and $\Psi_{iR}(x,y)$ for each flavor $i$ is given by
\cite{Gherghetta:2000qt}
\begin{equation}
\label{5dimenyc}
S_y=\int d^4x \int dy\, \sqrt{-g}\,\,\lambda_{ij(5d)} H(x)
   \Big( \bar\Psi_{iL}(x,y)\Psi_{jR}(x,y) + {\rm h.c.} \Big)
\delta(y-\pi R) \, .
\end{equation}
For simplicity we ignore flavor mixing, and further assume $c_{iL} = c_{iR} =
c_i$.  The Yukawa coupling of the zero mode fermions turns out to be
\cite{Gherghetta:2000qt}
\begin{equation}
\label{yc}
    \lambda_{i}= \lambda_{i(5d)}k (1/2-c_i)
\left(1- e^{(2c_i-1)\pi kR} \right)^{-1} \, .
\end{equation}
Assuming the 5d coupling $\lambda_{i(5d)} k \sim 1$, one can trade the zero
mode fermion masses in favor of the corresponding $c_i$ ($c_q = 0.62, 0.61,
0.51, 0.56, -0.49, 0.48$ for $q= u, d, c, s, t, b$). This is how the fermion
mass hierarchy problem is addressed.  Next, we derive the Yukawa coupling of
the $n$th KK fermion for $m_n \ll k \sim M_P$, $kR \gg 1$ and
${\lambda}_{i(5d)}k \sim 1$ as (with a tacit assumption of KK number
conservation to avoid any divergence in KK sum)
\begin{equation}
  \label{kkyukawa}
  {\lambda}_{i}^{(n)} \sim 
{\cos}^2 \Big(\left[ n-\frac{|c-\frac{1}{2}| -|c \mp \frac{1}{2}|}{2} - 
\frac{1}{2} \right] \pi \Big) \, ,
\end{equation}
where $\mp$ correspond to $\mathbb{Z}_2$ odd/even KK modes. Thus the KK Yukawa
couplings for $\mathbb{Z}_2$ even KK modes, regardless of their flavors and KK
numbers, are roughly equal to unity for the values of $c_q$ quoted above.

{\bf Contribution of KK states to ${\sigma(gg \to h)}$}:~ The process
$gg \rightarrow h$ proceeds through fermion triangle loops. The SM expression
of the cross section is given by ($\tau_q \equiv 4m_q^2 / m_{H}^2$)
\begin{eqnarray}
\label{ggh}
  \sigma_{gg \rightarrow h}^{\rm SM} & = &\frac{ \alpha_{s}^2}{576
    \pi v^2}\left|\sum_q A_q(\tau_q)\right|^2 \,\, , ~~{\rm where}~~ 
  \left. A_q(\tau_q) \right|_{\rm SM} = 2\tau_q[1+(1-\tau_q)f(\tau_q)] ~, 
 \\
\label{ftau}
  {\rm with}~ f(\tau) &=& {\rm arcsin}^2 \left( \frac{1}{\sqrt{\tau}}\right)
  ~{\rm for}~ \tau \geq 1, ~{\rm and}~
  f(\tau) = -\frac{1}{4} \left[ {\rm ln}\left( \frac{1+\sqrt{1-\tau}}
      {1-\sqrt{1-\tau}} \right) -i\pi \right]^2 {\rm for}~ \tau < 1 ~.
\end{eqnarray}
Above, $\alpha_s$ is the QCD coupling at the Higgs mass scale, $v$ is the
Higgs vacuum expectation value and $A_q$ is the loop amplitude from the $q$th
quark.  In the SM, the dominant contribution comes from the top quark loop.
Now, there will be additional contributions from the KK quarks. Importantly,
due to the large universal KK Yukawa couplings, not only the KK top but also
the KK modes of other quarks would have sizable contribution. Indeed, the
lightest modes ($n=1$) would have dominant contributions. Setting the KK
Yukawa couplings to unity, as suggested by Eq.~(\ref{kkyukawa}), we derive the
amplitude of the $n$th KK mediation of the $q$th flavor, with the same
normalization of Eq.~(\ref{ggh}), as
\begin{equation}
\left. A_q(\tau_{q_n}) \right|_{\rm KK} = \frac{4v^2}{m_{h}^2}
\left[1+(1-\tau_{q_n})f(\tau_{q_n})\right] . 
\label{FtKK}
\end{equation}
In 5d the sum over $n$ yields a finite result. Eq.~(\ref{FtKK}) is different
from the UED result \cite{Petriello:2002uu} in two ways: (i) we have set the
KK Yukawa coupling to unity irrespective of quark flavors, while in UED the KK
Yukawa coupling is controlled by zero mode masses; (ii) in UED there is an
additional factor of 2 because both $\mathbb{Z}_2$ even and odd KK modes
contribute, while in RS the odd modes do not couple to the brane-localized
Higgs.  In Fig.~1, we have plotted the variation with $m_h$ of the deviation
of the production cross section $\sigma_{\rm RS} (gg \to h)$ from its SM
expectation $\sigma_{\rm SM} (gg \to h)$ normalized by the SM value. The
dominant QCD correction cancels in this normalization. We have chosen four
reference values of $m_{\rm KK}$ ($= 1.0, 1.5, 2.0$ and $3.0$ TeV), where
$m_{\rm KK}$ is the KK mass of the $n=1$ gauge bosons, which also happens to
be the lightest KK mass in the bulk (corresponding to the conformal limit,
$c=1/2$ for fermions). For $m_h$ below 150 GeV, the deviation is quite
substantial (close to 45\%) for $m_{\rm KK} = 1$ TeV. For larger $m_{\rm KK}
=$ 1.5 (3.0) TeV, the effect is still recognizable, around 18\% (5\%). In the
inset, we exhibit a comparison between RS and UED contributions to the same
observable, where the KK mass scales of the two scenarios, namely $m_{\rm KK}$
for RS and $1/R$ for UED, have been assumed to be identical ($=1$ TeV). For
$m_h <$ 150 GeV, the RS contribution is about 2.5 times larger than the UED
contribution, while the margin slightly goes down with increasing $m_h$. This
factor 2.5 can be understood in the following way: In RS, five $n=1$ KK
flavors (except the KK top) have mass around $m_{\rm KK}$ with order one
Yukawa coupling. So naively we would expect a factor of 5 enhancement relative
to UED. But in UED both $\mathbb{Z}_2$ even and odd modes contribute. This
reduces the overall enhancement factor in RS over UED to about 2.5.

{\bf Contribution of KK states to ${\Gamma(h \to \gamma \gamma)}$}:~
The $h \to \gamma\gamma$ process proceeds through fermion triangles as
well as via gauge loops along with the associated ghosts. 
The decay width in the SM can be written as
\begin{equation}
\Gamma_{h \rightarrow \gamma \gamma} = \frac{ \alpha m_h^3}{256
 \pi^3 v^2}
\left| \sum_f N_c^f Q_f^2 A_f(\tau_f) + A_W(\tau_W)  
\right|^2 \, ,
\end{equation}
where $\alpha$ is the electromagnetic coupling at the Higgs mass scale.  The
expression for $A_f$ is given in Eq.~(\ref{ggh}), and the dominant SM
contribution to $A_f$ comes from the top quark loop.  The $W$-loop amplitude
in the SM is given by
\begin{equation}
\left. A_W (\tau_W)\right|_{\rm SM} = 
-\left[2+3\tau_W +3\tau_W(2-\tau_W)f(\tau_W)\right] \, .
\end{equation}
We derive the KK contribution of the gauge sector as 
\begin{equation}
\label{awn}
\left. A_W (\tau_{W_n})\right|_{\rm KK} =
-\left[2+3\tau_W +3\tau_W(2-\tau_{W_n})f(\tau_{W_n}) -
2(\tau_{W_n} -\tau_W)f(\tau_{W_n})\right]. 
\end{equation}
Again, the sum over $n$ yields finite result and in the limit of large KK mass
the KK contribution decouples. Our Eq.~(\ref{awn}) is very different from the
corresponding UED expression \cite{Petriello:2002uu}, primarily because the
Higgs is confined at the brane in the present scenario while it resides in the
bulk in UED.  In Fig.~2, we have plotted the decay width $\Gamma(h\to
\gamma\gamma)$ in RS relative (and normalized as well) to the SM. Again, the
four choices of $m_{\rm KK}$ are 1.0, 1.5, 2.0 and 3.0 TeV. There is a partial
cancellation between quark and gauge boson loops, both in real and imaginary
parts, not only for the zero mode but also for each KK mode. The meeting of
the four curves just above the $m_h = 2 m_t$ threshold is a consequence of the
above cancellation and at the meeting point the SM contribution overwhelms
the KK contribution. Unlike in Fig.~1, we witness both suppression and
enhancement with respect to the SM contribution. The inset carries an
illustration how RS fares against UED for identical KK masses.
\begin{figure*}
\begin{minipage}[t]{0.47\textwidth}
\begin{center}
\includegraphics[width=0.7\textwidth,angle=270,keepaspectratio]
{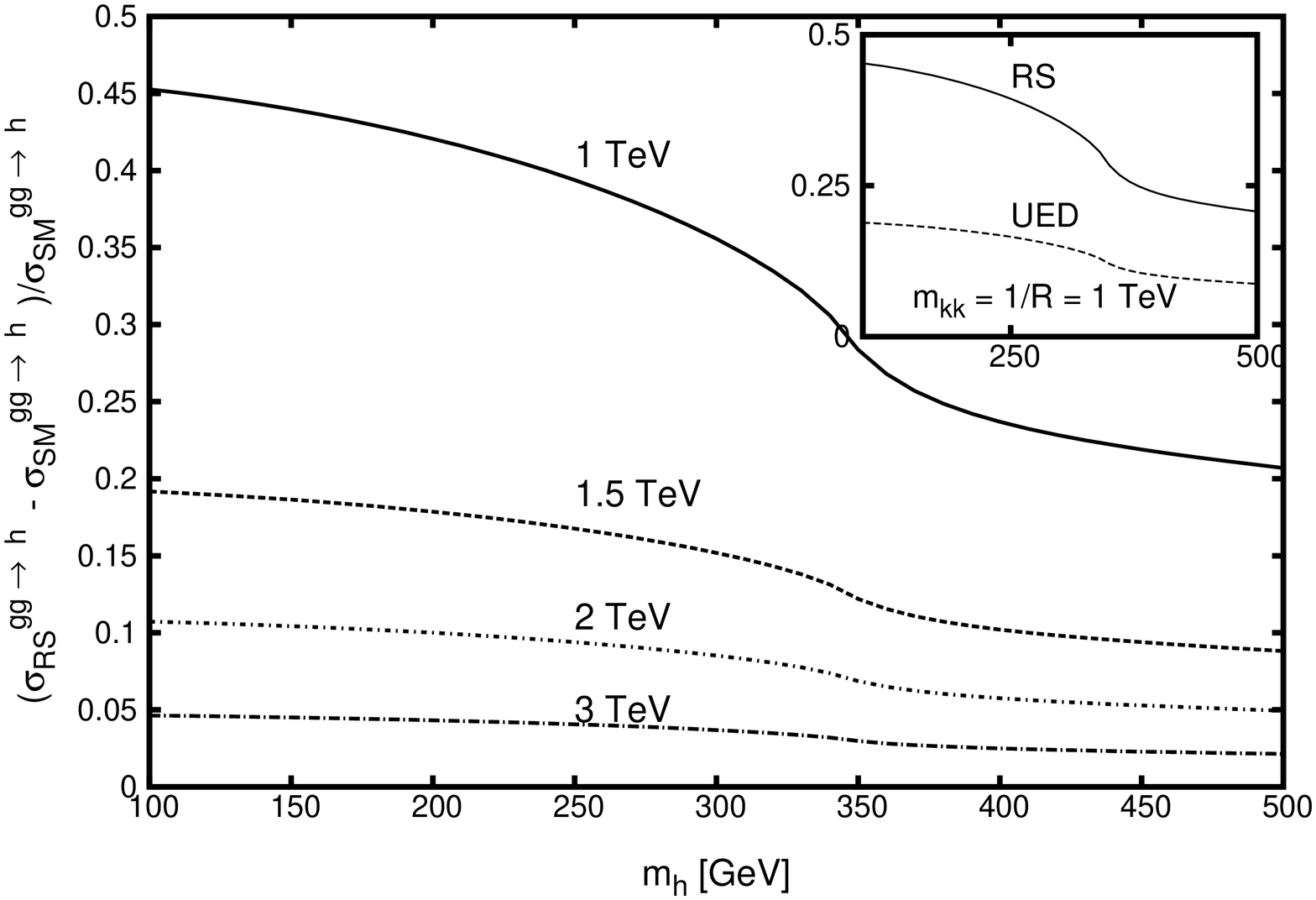}
\caption[]{\sf \small The fractional deviation (from the SM) of the $gg
  \rightarrow h$ production cross section in RS is plotted against the Higgs
  mass.  The four curves correspond to four different choices of $m_{\rm
    KK}$.  In the inset, we have compared the UED contribution for $1/R = 1$
  TeV with the RS contribution for $m_{\rm KK} = 1$ TeV.}
\end{center}
\end{minipage}
\hspace{7mm}
\begin{minipage}[t]{0.47\textwidth}
\begin{center}
\includegraphics[width=0.7\textwidth,angle=270,keepaspectratio]
{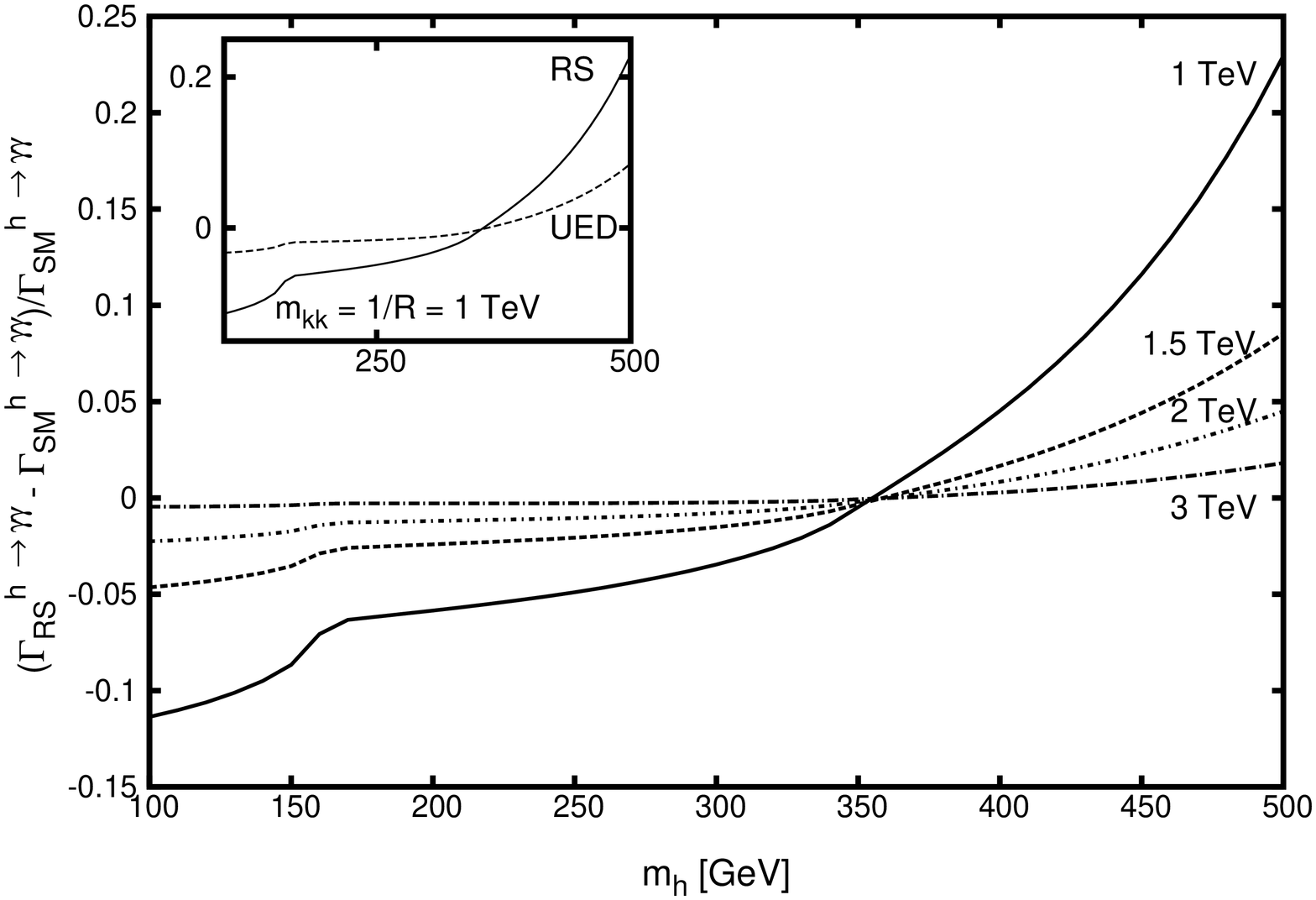}
\caption[]{\sf \small Same as in Fig.~1, except that the fractional deviation
  in $h \to \gamma \gamma$ decay width is plotted.}
\label{ggha}
\end{center}
\end{minipage}
\end{figure*}

Next we construct a variable $R = \sigma_{gg \to h}~\Gamma_{h \to \gamma
  \gamma}$. In Fig.~3, we have studied variation of $(R_{\rm RS} - R_{\rm
  SM})/R_{\rm SM}$ with $m_h$. For $m_{\rm KK} =$ 1.0, 1.5, 2.0 and 3.0 TeV,
the fractional changes in $R$ are 30\%, 14\%, 8\% and 4\%, respectively, for
$m_h < 150$ GeV. The comparison shown in the inset shows that RS wins over UED
roughly by a factor of 2 for identical KK scale for $m_h < 150$ GeV.
Incidentally, our UED plots in the insets of all the three figures are in
complete agreement with \cite{Petriello:2002uu}. See also \cite{Rai:2005vy}
for a numerical simulation of the Higgs signal at LHC in the UED context. 
\begin{figure}
\begin{center}
\includegraphics[width=0.4\textwidth,angle=270,keepaspectratio]
{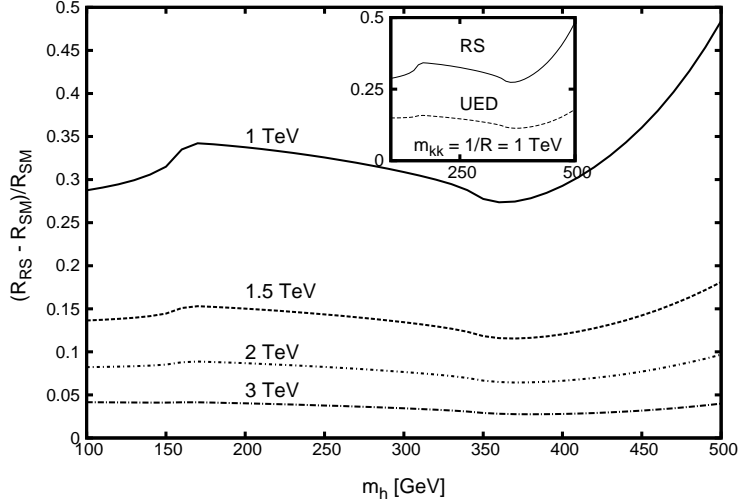}
\caption[]{\sf \small Same as in Fig.~1, except that the fractional deviation
  in $R = \sigma_{gg \to h}~ \Gamma_{h \to \gamma \gamma}$ has been plotted.}
\label{hgg}
\end{center}
\end{figure}

{\bf Conclusions}:~ In conclusion, we highlight the core issues: In the RS
scenario, the brane-bound Higgs can have order one Yukawa coupling with the KK
fermions of all flavors. Such large KK Yukawa couplings can sizably enhance
the Higgs production via gluon fusion and alter the Higgs decay width into two
photons, provided the KK masses are in a regime accessible to the LHC. Because
of the proactive involvement of more flavors inside the loop the effect in RS
is significantly stronger (typically, by a factor of 2 to 2.5) than in UED for
similar KK masses. Admittedly, this advantage in RS is somewhat offset by the
fact that the lightest KK mass in UED can be as low as 500 GeV thanks to the
KK-parity, while in RS a KK mass below 1.5 TeV would be difficult to
accommodate (see below). However, attempts have been made to impose KK parity
in warped cases as well \cite{Agashe:2007jb}.

Electroweak precision tests put a severe lower bound on $m_{\rm KK}$ ($\sim
10$ TeV) \cite{Hewett:2002fe}. To suppress excessive contribution to $T$ and
$S$ parameters the gauge symmetry in the bulk is extended to ${\rm SU(2)_L
  \times SU(2)_R \times U(1)_{B-L}}$, and then $m_{\rm KK}$ as low as 3 TeV
can be allowed \cite{Agashe:2003zs,Bouchart:2008vp}. A further discrete
symmetry $L \to R$ helps to suppress $Zb_L\bar b_L$ vertex correction and
admits an even lower $m_{\rm KK} \sim 1.5$ TeV \cite{Carena:2007ua}.  If some
other new physics (e.g. supersymmetrization of RS) can create further room in
$T$ and $S$ by partial cancellation, $m_{\rm KK} \sim$ 1 TeV can also be
accommodated.  In our analysis, values of $m_{\rm KK}$ in the range of 1-3 TeV
chosen for illustration may arise in the backdrop of such extended symmetries.
Furthermore, if the $b'$ quark, present in the case of left-right gauge
symmetry, weighs around 1 TeV, one obtains an {\em additional} $\sim$ 10\%
contribution to $\sigma(gg \to h)$ \cite{Djouadi:2007fm}. 

A very recent paper \cite{Cacciapaglia:2009ky} lists the relative contribution
of different scenarios (supersymmetry, flat and warped extra dimension, little
Higgs, gauge-Higgs unification, fourth generation, etc.)  to $gg\to h$ and $h
\to \gamma\gamma$ for some benchmark values.  A comparison between their work
and ours in order.  As regards the RS scenario, the authors of
\cite{Cacciapaglia:2009ky} consider the region of parameters where the zero
mode quarks mix with their KK partners. Additionally, their choice of $c_L$ is
substantially different from $c_R$, where they observe large destructive
interference in the effective $ggh$ coupling.  On the other hand, our working
hypothesis is based on: $c \equiv c_L = c_R$ (see Eq.~(\ref{yc})), and we
assume KK number conservation at the Higgs vertex.  We observe that the Higgs
coupling to KK quarks is large for any flavor (see Eq.~(\ref{kkyukawa})), and
the (direct) loop effects of the KK quarks (which carry the same quantum
numbers as their zero modes) do enhance the effective $ggh$ vertex (like the
{\em enhancement} observed for the fourth family contribution
\cite{Cacciapaglia:2009ky}, or the $b'$ quark contribution
\cite{Djouadi:2007fm}, or the UED contribution
\cite{Petriello:2002uu,Cacciapaglia:2009ky}), and the magnitude is rather
insensitive to the value of $c$ as long as $|c| ~\gtap ~ 0.5$.  The authors of
\cite{ghu} also calculate the KK-induced effective $ggh$ vertex, but they rely
on the gauge-Higgs unification set-up, and hence an efficient numerical
comparison of their work with ours is not possible.

\noindent{ \bf Acknowledgments:}
TSR acknowledges the support (S.P. Mukherjee fellowship) of CSIR, India. GB's
research is partially supported by project no.~2007/37/9/BRNS (DAE), India. 
We thank G. Cacciapaglia for making valuable comments.

\end{document}